\documentstyle[12pt]{article}
\input psfig.sty
\def\nn{\noindent}
\def\Re{{\cal R \mskip-4mu \lower.1ex \hbox{\it e}\,}}
\def\Im{{\cal I \mskip-5mu \lower.1ex \hbox{\it m}\,}}
\def\ie{{\it i.e.}}
\def\eg{{\it e.g.}}
\def\etc{{\it etc}}
\def\etal{{\it et al.}}

\def\sub#1{_{\lower.25ex\hbox{$\scriptstyle#1$}}}
\def\tev{\,{\ifmmode\mathrm {TeV}\else TeV\fi}}
\def\gev{\,{\ifmmode\mathrm {GeV}\else GeV\fi}}
\def\mev{\,{\ifmmode\mathrm {MeV}\else MeV\fi}}
\def\mpl{\ifmmode \overline M_{Pl}\else $\overline M_{Pl}$\fi}
\def\to{\rightarrow}

\def\subw{_{\rm w}}
\def\mh{\ifmmode m\sbl H \else $m\sbl H$\fi}
\def\mch{\ifmmode m_{H^\pm} \else $m_{H^\pm}$\fi}
\def\mt{\ifmmode m_t\else $m_t$\fi}
\def\mc{\ifmmode m_c\else $m_c$\fi}
\def\mz{\ifmmode M_Z\else $M_Z$\fi}
\def\mw{\ifmmode M_W\else $M_W$\fi}
\def\mws{\ifmmode M_W^2 \else $M_W^2$\fi}
\def\mhs{\ifmmode m_H^2 \else $m_H^2$\fi}   
\def\mzs{\ifmmode M_Z^2 \else $M_Z^2$\fi}
\def\mts{\ifmmode m_t^2 \else $m_t^2$\fi}
\def\mcs{\ifmmode m_c^2 \else $m_c^2$\fi}
\def\mchs{\ifmmode m_{H^\pm}^2 \else $m_{H^\pm}^2$\fi}
\def\ztwo{\ifmmode Z_2\else $Z_2$\fi}
\def\zone{\ifmmode Z_1\else $Z_1$\fi}
\def\mtwo{\ifmmode M_2\else $M_2$\fi}
\def\mone{\ifmmode M_1\else $M_1$\fi}
\def\tb{\ifmmode \tan\beta \else $\tan\beta$\fi}
\def\xw{\ifmmode x\subw\else $x\subw$\fi}
\def\ch{\ifmmode H^\pm \else $H^\pm$\fi}
\def\lum{\ifmmode {\cal L}\else ${\cal L}$\fi}
\def\inpb{\,{\ifmmode {\mathrm {pb}}^{-1}\else ${\mathrm {pb}}^{-1}$\fi}}
\def\infb{\,{\ifmmode {\mathrm {fb}}^{-1}\else ${\mathrm {fb}}^{-1}$\fi}}
\def\epem{\ifmmode e^+e^-\else $e^+e^-$\fi}
\def\ppb{\ifmmode \bar pp\else $\bar pp$\fi}
\def\bsg{\ifmmode B\to X_s\gamma\else $B\to X_s\gamma$\fi}
\def\bsll{\ifmmode B\to X_s\ell^+\ell^-\else $B\to X_s\ell^+\ell^-$\fi}
\def\bstt{\ifmmode B\to X_s\tau^+\tau^-\else $B\to X_s\tau^+\tau^-$\fi}
\def\lamt{\ifmmode \tilde\lambda\else $\tilde\lambda$\fi}
\def\shat{\ifmmode \hat s\else $\hat s$\fi}
\def\that{\ifmmode \hat t\else $\hat t$\fi}
\def\uhat{\ifmmode \hat u\else $\hat u$\fi}

\newskip\zatskip \zatskip=0pt plus0pt minus0pt
\def\matth{\mathsurround=0pt}

\def\gsim{\mathrel{\mathpalette\atversim>}}
\def\atversim#1#2{\lower0.7ex\vbox{\baselineskip\zatskip\lineskip\zatskip
  \lineskiplimit 0pt\ialign{$\matth#1\hfil##\hfil$\crcr#2\crcr\sim\crcr}}}

\renewcommand{\thefootnote}{\fnsymbol{footnote}}

\begin{document} \begin{titlepage} 
\rightline{\vbox{\halign{&#\hfil\cr
%&DRAFT\cr
&SLAC-PUB-8746\cr
&January 2001\cr}}}
\begin{center}

{\Large\bf Cartography with Accelerators: Locating Fermions in Extra 
Dimensions at Future Lepton Colliders}
\footnote{Work supported by the Department of 
Energy, Contract DE-AC03-76SF00515}
\medskip

\normalsize 
{\bf \large Thomas G. Rizzo}
\vskip .3cm
Stanford Linear Accelerator Center \\
Stanford University \\
Stanford CA 94309, USA\\
\vskip .2cm

\end{center}

\begin{abstract} 
In the model of Arkani-Hamed and Schmaltz the various chiral fermions of the 
Standard Model(SM) are localized at different points on a thick wall which 
forms an extra dimension. Such a scenario provides a way of understanding the 
absence  of proton decay and the fermion mass hierarchy in models with extra 
dimensions. In this paper we explore the capability of future lepton colliders 
to determine the location of these fermions in the extra dimension through 
precision measurements of conventional scattering processes both below and on 
top of the lowest lying Kaluza-Klein gauge boson resonance. We show that for 
some classes of models the locations of these fermions can be very precisely 
determined while in others only their relative positions can be well measured. 
\end{abstract} 

%\vskip0.45in
%\begin{center}

%Submitted to Physical Review {\bf D}.

%\end{center}

\renewcommand{\thefootnote}{\arabic{footnote}} \end{titlepage}

%%%%%%%%%%%%%%%%%%%%%%%%%%%%%%%---- Put text here

\section{Introduction}

The possibility that the gauge bosons of the Standard Model(SM) may be 
sensitive to the existence of extra dimensions near the TeV scale has been 
entertained for some time; in the literature the cases with both 
factorizable{\cite{antoniadis}} and non-factorizable{\cite {us1}} metrics(such 
as in the Randall-Sundrum(RS) model{\cite {RS}}) have been considered. The 
phenomenology of these models in either case is particularly sensitive to how 
the SM fermions are treated. 

In the simplest scenario, the fermions remain on 
the wall located at the fixed point $y_i=0$ and are not free to experience 
the extra dimensions. (Here, lower-case Roman indices 
label the co-ordinates of the additional dimensions while Greek indices 
label our usual 4-d space-time.) However, since 5-d translational invariance is 
broken by the wall, the SM fermions interact with the Kaluza-Klein(KK) 
tower excitations of the SM gauge fields in the usual trilinear manner, \ie, 
$\sim gC_n\bar f \gamma_\mu f G^\mu_{(n)}$, with the $C_n$ being some 
geometric factor and $n$ labelling the KK tower state with which the fermion 
is interacting. For simple factorizable models, current low-energy constraints 
arising 
from, \eg, $Z$-pole data, the $W$ boson mass and $\mu$-decay generally require 
the mass of the lightest KK gauge boson to be rather heavy,  $\gsim 4$ TeV in 
the case of the 5-d SM{\cite {bunch}}. In the case of the RS model, similar 
analyses have obtained lower bounds as great as 25 TeV{\cite {us1}}. 

A second 
possibility, perhaps the most democratic, allows all of the SM fields to 
propagate in the $\sim$ TeV$^{-1}$ bulk{\cite {ACD}}. In this case, there being 
no matter on the 
walls, the conservation of momentum in the extra dimensions is restored and 
one now obtains interactions in the 4-d Lagrangian of the form 
$\sim gC_{ijk}\bar f^{(i)} \gamma_\mu f^{(j)} G^\mu_{(k)}$, which for 
factorizable metrics vanishes unless $i+j+k=0$, as a result of the afore 
mentioned momentum conservation. This implies that pairs of 
zero-mode fermions, which we identify with those of the SM, cannot directly 
interact singly with any of the excited modes in the gauge boson KK towers.  
Such a situation  
clearly limits any constraints arising from precision measurements since zero 
mode fermion fields can only interact with pairs of tower gauge boson fields. 
In addition, at colliders it now follows that 
KK states must be pair produced, thus significantly 
reducing the possible direct search reaches for these states. In fact, 
employing constraints from current experimental data, Appelquist, Cheng and 
Dobrescu(ACD){\cite {ACD}} find that the KK states in this 
scenario can be as light as 
a few hundred GeV. In the RS case, since the bulk is now AdS$_5$, 
the wave functions for the zero-mode fermions are no longer 
$y$-independent and the higher KK modes are described by Bessel 
functions, the existence of non-zero $C_{ijk}$ does 
not require the above index sum to vanish. However, one still finds that the 
low-energy constraints on the gauge boson KK mass scale in this 
scenario are also weaker than when fermions are forced to lie on the wall---but 
not to the degree experienced by the factorizable models{\cite {us2}}. 

Perhaps the most interesting possibility occurs when the SM fermions experience 
extra dimensions by being `stuck', \ie, localized or trapped at different 
specific points in a thick brane{\cite {nam}} away from the conventional fixed 
points. It has been shown that such a scenario can explain the absence of a 
number of rare processes, such as proton decay, by geometrically 
suppressing the size of the Yukawa couplings associated with the relevant 
higher dimensional operators without resorting to the existence of additional 
symmetries of any kind. In addition such a scenario may be able to explain the 
fermion mass hierarchy and the 
observed CKM mixing structure thus addressing important issues in flavor 
physics{\cite {mirsch}}. In order for this scheme to work the field that traps 
the various fermions must make the width of their wave functions in the extra 
dimension, $\sigma$, both much smaller than the typical separation of the 
various fermions as well as the size of the extra dimensions themselves 
$\sim \pi R_c$, where $R_c$ is the compactification radius. In the original 
model of Arkani-Hamed and Schmaltz the fermion wave functions were taken to 
be rather 
narrow Gaussians, \eg, $\sigma \sim {1\over {10}} R_c$. Previous analyses
{\cite {nym,me}} have indicated that it may be possible to obtain some 
information about the specific locations of the various SM fermions in the 
extra dimensions at future colliders through measurements of cross sections 
and various asymmetries. In particular it has been demonstrated that future 
lepton colliders can distinguish the case where quarks and leptons are 
localized at the same fixed point from where they are separated by a distance 
$\pi R_c${\cite {me}}, \ie, they sit at opposite fixed points. 
In this paper we will address this issue in detail, \ie, 
whether and/or how well future lepton colliders can be used to determine the 
locations of all of the SM fermion zero-modes in the extra dimensions.

The outline of this paper is as follows: in Section II we describe our 
setup in detail and will employ the 
current precision electroweak data to obtain lower bounds on the masses of the 
first KK gauge bosons in the Arkani-Hamed-Schmaltz model. Such an analysis has 
yet to be performed for the case of models with localized fermions with the 
three generations localized at different points and as a first step is 
necessary to determine 
the energy scale that needs to be explored by future lepton colliders. 
In Section III we will demonstrate the ability of these colliders to determine 
the various fermion locations in the case of one extra dimension both with 
and without the added assumption of any family symmetry. We will consider 
measurements both below and on top of the first gauge KK resonance; we will 
demonstrate that it is the below resonance measurements that are more directly 
useful in localizing the SM fermions. This is particularly useful if the center 
of mass energy of the lepton collider were limited. Finally, our summary and 
conclusions can be found in Section IV.

\section{Setup and Bounds from Precision Data}

In the Arkani-Hamed-Schmaltz scenario{\cite {nam}}, a fermion (in particular 
the fermion zero mode) interacts with a new scalar field which generates the 
thick/fat brane or domain wall{\cite {domwall}} and which has 
a zero determined by its Yukawa coupling at the point at 
which the fermion is to be trapped. To be specific let us consider the case of 
one extra dimension. In the region near the zero the scalar field is 
essentially a linear function of $y$ thus leading to an approximately 
Gaussian-shaped 
wave function for the trapped fermion. This region of linearity is rather 
small, $\sim \sigma$, in comparison to size of the compactified space $\sim 
\pi R_c$, and its slope in that region sets the scale for the Gaussian width 
of the fermion wave function. Unlike the fermions, the SM gauge and 
Higgs fields are free to propagate throughout the brane thickness and 
it is the zero mode of the Higgs which obtains the usual SM 
vacuum expectation value(vev) which is therefore $y$ independent. Choosing 
as usual the $S^1/Z_2$ 
orbifold for compactification, the gauge fields can be straightforwardly 
decomposed into their KK modes(Case I) and we are also free to choose the 
gauge where the fifth component of the gauge fields vanishes; thus
\begin{equation}
G_\mu(x,y)=G_\mu^{(0)}(x)+\sqrt 2 \sum_{n=1} \Bigg[G_\mu^{+(n)}(x)
\cos {ny\over {R_c}}+G_\mu^{-(n)}(x)\sin {ny\over {R_c}}\Bigg]\,,
\end{equation}
with the $\pm$ referring to the $Z_2$ parity of the KK tower states. 
Representing the $y$-dependent part of the wave function for the fermion zero 
modes as a set of 
normalized Gaussians with a common width $\sigma$, $g(y-y_f)$, centered 
around the points $y_f$, the interaction of the gauge and fermion fields can be 
described by the action 
\begin{equation}
S=\int ~d^4x dy ~g_5\bar f(x)g(y-y_f)\gamma_\mu f(x)g(y-y_f)G^\mu(x,y)\,,
\end{equation}
so that the gauge zero modes, the usual SM gauge fields, have couplings which 
are insensitive to the fermions position, $y_f$. In the limit of very narrow 
Gaussians, $(n\sigma/\pi R_c)^2 <<1$, the gauge KK excitations do not see the 
finite width of the fermions in the extra dimension and we are essentially 
left in the $\delta$-function 
limit. In that case the $y$ integration is trivial and we find that 
the gauge KK tower modes interact with the fermion zero modes via the action 
\begin{equation}
S=\int ~d^4x ~{\sqrt 2} g_4\bar f(x)\gamma^\mu f(x)\sum_{n=1} 
\Bigg[G_\mu^{+(n)}(x)
\cos {ny_f\over {R_c}}+G_\mu^{-(n)}(x)\sin {ny_f\over {R_c}}\Bigg]\,.
\end{equation}
As discussed in Refs. {\cite {nym,me,domwall}}, when $n$ grows sufficiently 
large the above 
inequality is no longer satisfied and the gauge fields begin to see the 
finite size of the fermion wave function and the resulting integrated overlap 
of the fermion and gauge wave functions results in an exponential suppression 
of the coupling of the fermion zero modes to the higher KK gauge boson tower 
members. (Further effects due to the fact that the wave functions are not 
truely Gaussian in shape will also occur.) This is 
particularly important in the case of more than one extra 
dimension where sums over KK states are generally divergent. This exponential 
suppression of the couplings for large $n$ now results in convergent sums 
over KK states for these cases. 

As an aside we note that in the above KK decomposition we have not demanded 
that the gauge field itself have a fixed $Z_2$ parity. In many cases this is 
useful in model construction and, since the SM fields are to be identified 
with the zero modes, gauge fields are taken to be $Z_2$ even. In this 
circumstance, which we call case II, the KK decomposition is identical 
to the above after dropping the terms 
proportional to $\sin {ny_f\over {R_c}}$.

Since the Higgs field obtains its constant vev in the bulk there is 
no tree level mixing between the different 
gauge KK tower levels and after spontaneous symmetry 
breaking (SSB) it is easiest to employ the conventional SM/mass eigenstate 
basis for the tower 
states: $\gamma^{\pm (n)}, Z^{\pm (n)}, W^{\pm (n)}$, \etc. While the 
$\gamma^{\pm (n)}$ obtain masses $n/R_c=nM_c$, the $Z^{\pm (n)}$ and 
$W^{\pm (n)}$ states have masses given by $(n^2M_c^2+M_{Z,W}^2)^{1/2}$. This 
implies that, \eg, if $M_c$=4 TeV then the difference in mass between the 
first $Z$ and $\gamma$ KK excitation is approximately 1 GeV. In the case where 
both $Z_2$ even and odd gauge KK states are present, the two are exactly mass 
degenerate at the tree level. In the absence of tree-level mixing of any of 
the KK states, their modifications to the precision electroweak parameters are 
relatively straightforward to isolate and in doing so we follow the analysis 
as presented by Rizzo and Wells{\cite {bunch}}.

First, we assign the 15 SM chiral fermions specific locations in the extra 
dimension: $y_{Q_i}, y_{L_i}, y_{u_i}, y_{d_i}$ and $y_{e_i}$ where 
$Q_i^T=(u,d)_{Li}$, $L^T=(\nu, \ell)_{Li}$, \etc. These locations 
are found to be somewhat more easily 
analyzed in terms of the dimensionless rescaled variables $x_a=y_a/\pi R_c$. 
Given the fact that the KK gauge boson mass matrix is diagonal, the first 
place that KK tower exchange will be important is in $\mu$-decay through which 
the Fermi constant is defined. At tree level one can write 
\begin{equation}
{G_F\over {\sqrt 2}}={g^2\over {8M_W^2}}(1+V_G)\,,
\end{equation}
where $V_G$ sums the W boson tower KK contributions; SM radiative corrections 
to this expression can be performed in the usual manner and we will assume 
these are included in what follows. (We note in passing that, in principle, 
stronger bounds on the scale $M_c$ can be obtained by considering constraints 
on leptonic flavor-changing neutral currents(FCNC) as discussed by Delgado, 
Pomarol and Quiros{\cite {bunch}}. However, in our particular case these 
bounds will 
involve all six of the coefficients $x_{L_i}$ and $x_{e_i}$ as well as 
the {\it a priori} unknown left and right-handed leptonic mixing matrices. 
This makes it somewhat difficult to extract any useful limits without making 
further model-dependent assumptions. Of 
course, if the $x$'s were generation independent, FCNC would be avoided at 
tree level as in the SM. For the moment, however, we will attempt 
to remain as general as possible.) In the 5-d SM 
example above, neglecting the possible effects of the Gaussian coupling 
suppression as previously discussed,  we can write
\begin{equation}
V_G={2M_W^2\over {M_c^2}}\sum_{n=1}{1\over {n^2}}(\cos \pi nx_{L_1}
\cos \pi nx_{L_2}+\sin \pi nx_{L_1} \sin \pi nx_{L_2})\,,
\end{equation}
with $x_{L_{1,2}}$ being the locations of the left-handed lepton doublets of 
the first two generations. In the case where gauge fields are $Z_2$ odd the 
last term can be dropped otherwise the two terms can be combined into a single 
term : $\cos \pi n(x_{L_1}-x_{L_2})$, which means that the correction due to
tower exchanges depends only upon the relative positions of the two lepton 
doublets. In this case we find that we can write $V_G$ as 
\begin{equation}
V_G={\pi^2 M_W^2\over {3M_c^2}}F(\Delta)\,,
\end{equation}
with $\Delta=x_{L_1}-x_{L_2}$ and $F(\Delta)$ shown in Fig.1. With the help of 
Ref. {\cite {nym}} one finds that $F(\Delta)$ can be obtained analytically: 
$F(\Delta)=1-3\Delta+{3\over {2}}\Delta^2$. Note that in the limit where the 
$x$'s are generation independent, $\Delta=0$ so that $F=1$ and cases I and II 
yield identical results. It is very 
important to observe that,  
unlike the parameter $V$ introduced into the Rizzo and Wells{\cite {bunch}} 
analysis, $V_G$ can have either sign and even vanishes when $\Delta \simeq 
0.423$. In the case where the $Z_2$ odd pieces are dropped we can still 
re-write $V_G$ into a form similar to the above with the replacement 
$F(\Delta)\to \tilde 
F(x_{L_1},x_{L_2})$ which remains a function of two variables. Scanning over 
the $x_{L_i}$, however, we again find that the sum, \ie, $\tilde F$ is bounded 
to the region $-{1\over {2}}\leq \tilde F \leq 1$ as was $F(\Delta)$. 

\vspace*{-0.5cm}
\nn
\begin{figure}[htbp]
\centerline{
\psfig{figure=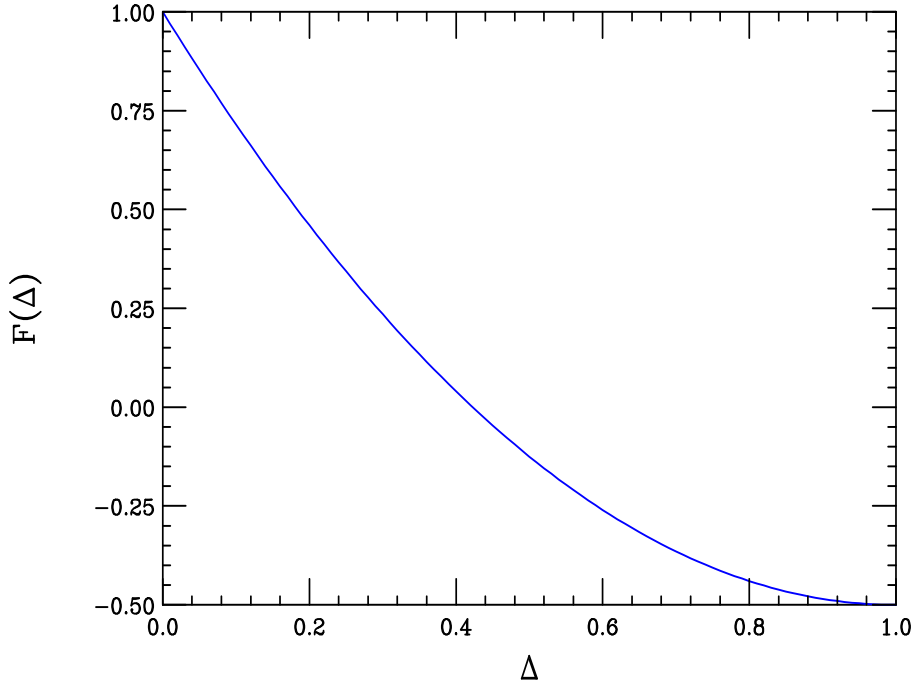,height=9cm,width=12cm,angle=90}}
\vspace*{0.5cm}
\caption[*]{The function $F(\Delta)$.}
\label{lims}
\end{figure}
\vspace*{0.4mm}

Given the shift in $G_F$ due to KK exchange it is straightforward to calculate 
the first order shifts in other electroweak observables due to $V_G$ which we 
imagine (due to the great success of the SM) to be rather small, \ie, of 
order $10^{-3}$ or less. Thus we are free in our analysis below to drop terms 
in $V_G$ that are of more than linear order; this assumption is justified by 
our final results. We find that

\begin{eqnarray}
\delta x_0 &=& {x_0(1-x_0)\over {(1-2x_0)}}V_G\nonumber \\
\delta M_W &=& -{1\over {2}}M_W {x_0\over {(1-2x_0)}}V_G\nonumber \\
{\delta \Gamma_f\over {\Gamma_f}} &=& -\Bigg[1+{8Q_f(2T_{3f}-4x_0Q_f)\over 
{1+(2T_{3f}-4x_0Q_f)^2}} {x_0(1-x_0)\over {(1-2x_0)}}\Bigg]V_G\,,
\label{basic}
\end{eqnarray}
where $x_0(M_W)$ is the value of the effective weak mixing angle, 
$\sin ^2 \theta_{eff}$ ($W$ boson 
mass) and $\Gamma_f$ is the decay width for $Z\to \bar f f$ in the SM 
after all of the SM QED, QCD and electroweak radiative corrections 
have been included. Here $T_{3f}$ and $Q_f$ are the third component of weak 
isospin and the electric charge of the fermion, $f$, respectively. 
For $f=b,c$, the widths are 
usually quoted through the ratios $R_f=\Gamma(Z\to \bar f f)/
\Gamma(Z\to hadrons)$; the shifts in these ratios are given by 
\begin{equation}
{\delta R_i\over {R_i}}={\delta \Gamma_i\over {\Gamma_i}}-
{\sum_j \delta  \Gamma_j\over {\sum_j \Gamma_j}}\,,
\end{equation}
where both sums in the last expression extend over all hadrons. 
These are the electroweak variables we will employ in our fit. Unlike in the 
case of Rizzo 
and Wells, we cannot make use of other observables, such as weak charge $Q_W$ 
measured in atomic parity violation or data arising from 
deep inelastic neutrino scattering, since they necessarily 
involve five of the $x_i$ for the neutrino/electron and the $u$ and $d$ quarks.

\vspace*{-0.5cm}
\nn
\begin{figure}[htbp]
\centerline{
\psfig{figure=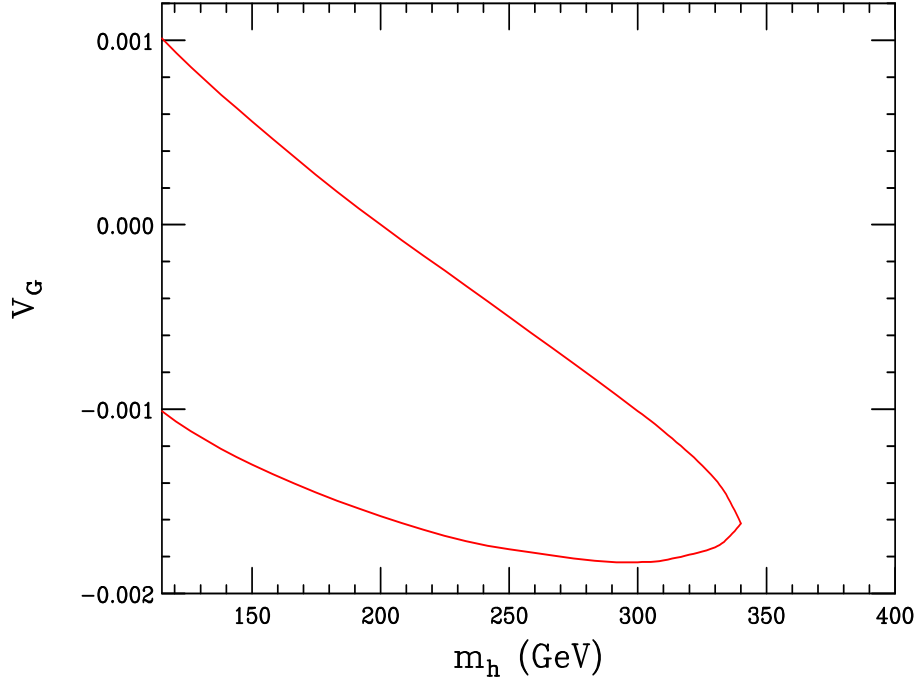,height=9cm,width=12cm,angle=90}}
\vspace*{15mm}
\centerline{
\psfig{figure=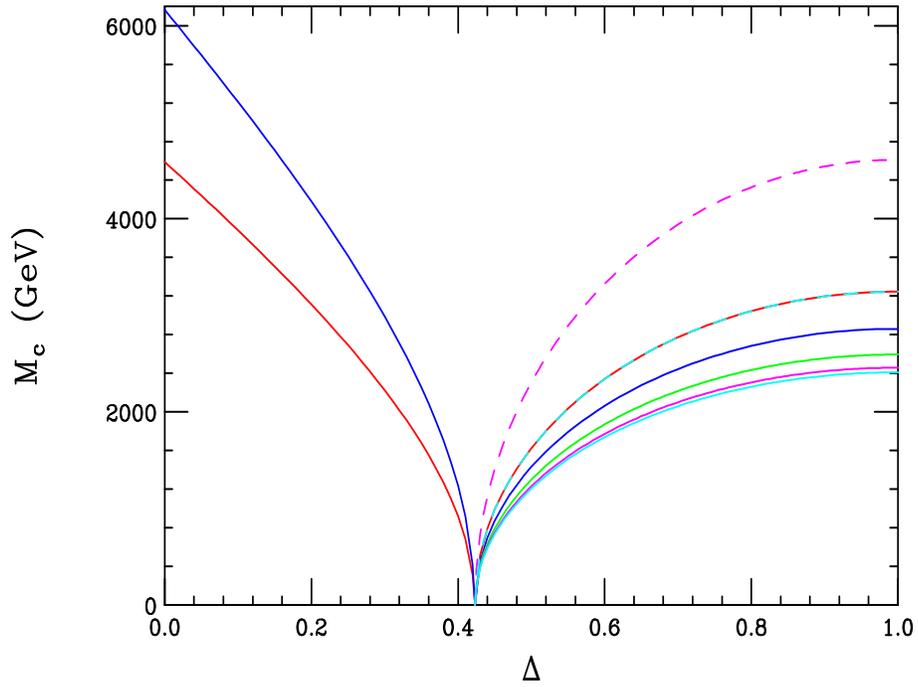,height=9cm,width=12cm,angle=90}}
%\vspace*{-0.9cm}
\caption{(top)$95\%$ CL allowed region in the $V_G-m_h$ plane resulting from a 
fit to electroweak data. (bottom) $95\%$ CL mass bounds as a function of 
$\Delta$ resulting from the electroweak fit. The solid red(blue,green) 
corresponds to the lower bound for $m_h=115(150,200)$ GeV. For $m_h=250(300)$ 
GeV the purple(cyan) solid and dashed pair of curves give the lower and upper 
bounds on $M_c$.}
\label{fun}
\end{figure}
\vspace*{0.4mm}

Using the latest precision electroweak 
data{\cite {precdata}} as well as the updated estimate of 
$\alpha(M_Z)${\cite {alpha}} obtained from the new low energy data on 
$R${\cite {Rdata}} from the BES-II Collaboration, we find the $95\%$ CL allowed 
region in the $V_G$-Higgs mass($m_h$) plane shown in Fig.2 for Case I. (All 
electroweak radiative correction 
calculations are performed with ZFITTER6.23{\cite {Zfitter}}.) 
We recall that 
this same fit in the SM requires $m_h<200$ GeV at this level of confidence; 
here we see that the extra 
degree of freedom present in $V_G$ allows the SM Higgs mass to be as heavy as 
350 GeV. Note that for $m_h>200$ GeV the $95\%$ CL plot {\it requires} 
$V_G <0$. Since $F(\Delta)$ is known we can use these results to obtain a 
bound on $M_c$ as a function of $\Delta$ for fixed values of $m_h$ which is 
also shown in Fig.2 for Case I. The first thing to note is that in the case 
where the $x$'s are family independent, \ie, $\Delta=0$ the value of $M_c$ 
must be larger than about 4 TeV given the current lower bound on the Higgs 
mass $\simeq 113$ GeV. (This same result applies also for Case II.) We 
see that as we vary both $m_h$ and $\Delta$ the resulting bounds range from 
zero to over 6 TeV with typical values of order 2-5 TeV. A scan of the 
model parameters for Case II gives qualitatively very similar results. From 
this analysis it is clear that a bound of $M_c=4$ TeV is rather typical and we 
will use this value in our analysis below. To reach such KK resonance mass 
scales will either 
require CLIC{\cite {clic}} technology or a muon collider{\cite {muon}} either 
of which should have integrated luminosities in excess of 1 ab$^{-1}$=1000 
fb$^{-1}$. Thus it is of particular importance what we can learn from data 
available below the first KK peak.

\section{Localizing Fermions at Lepton Colliders}

\subsection{Bhabha Scattering}

In order to localize the various SM fermions we must first determine the 
positions of the particles in the initial state, \ie, for an 
$e^+e^-(\mu^+\mu^-)$ collider we must determine the values of $x_{L_{1(2)}}$ 
and $x_{e_{1(2)}}$. (From now on we will drop the family index except where 
such labels are important and we will denote $c_{L(R)}=\cos n\pi x_{L(e)}$ 
and $s_{L(R)}=\sin n\pi x_{L(e)}$.) This can best be done via Bhabha 
scattering for which the cross section is given by
\begin{eqnarray}
{d\sigma \over {d\cos \theta}} &=& {2\pi \alpha \over {s}} \sum_{i,j}
\Bigg[\bigg([LL](\ell_i\ell_j)^2+[RR](r_ir_j)^2\bigg) {u^2\over {s^2}}
(P_{ij}^{ss}+2P_{ij}^{st}+P_{ij}^{tt})\nonumber \\
&+& (\ell_i\ell_j)(r_ir_j)\bigg(([LL]+[RR]){t^2\over {s^2}}P_{ij}^{ss}+[LR]
P_{ij}^{tt}\bigg)\Bigg]\,,
\end{eqnarray}
where $\ell_i$ and $r_i$ are the couplings of the electron(or muon) to the 
various gauge bosons, \eg, $\ell_Z={g\over {\cos \theta_w}}(-{1\over {2}}+
\sin^2 \theta_w)$. Here we have employed the usual Mandelstam variables and 
have made use of the notation 
\begin{equation}
P_{ij}^{qr}=s^2 {{(q-m_i^2)(r-m_j^2)+\Gamma_i \Gamma_j m_i m_j}\over {
[(q-m_i^2)^2+(\Gamma_i m_i)^2][(r-m_j^2)^2+(\Gamma_j m_j)^2]}}\,,
\end{equation}
with $m_i (\Gamma_i)$ being the mass (width) of the $i^{th}$ gauge boson
and where we have defined the polarization projectors {\cite {cuypers}}, 
\begin{eqnarray}
{[LL]} &=& {1\over {4}}[1+P_1+P_2+P_1P_2]\,, \nonumber \\
{[RR]} &=& {1\over {4}}[1-P_1-P_2+P_1P_2]\,, \nonumber \\
{[LR]} &=& {1\over {2}}[1-P_1P_2]\,,
\end{eqnarray}
with $P_{1,2}$ being the polarizations of the incoming electron and positron 
beam respectively.

\vspace*{-0.5cm}
\nn
\begin{figure}[htbp]
\centerline{
\psfig{figure=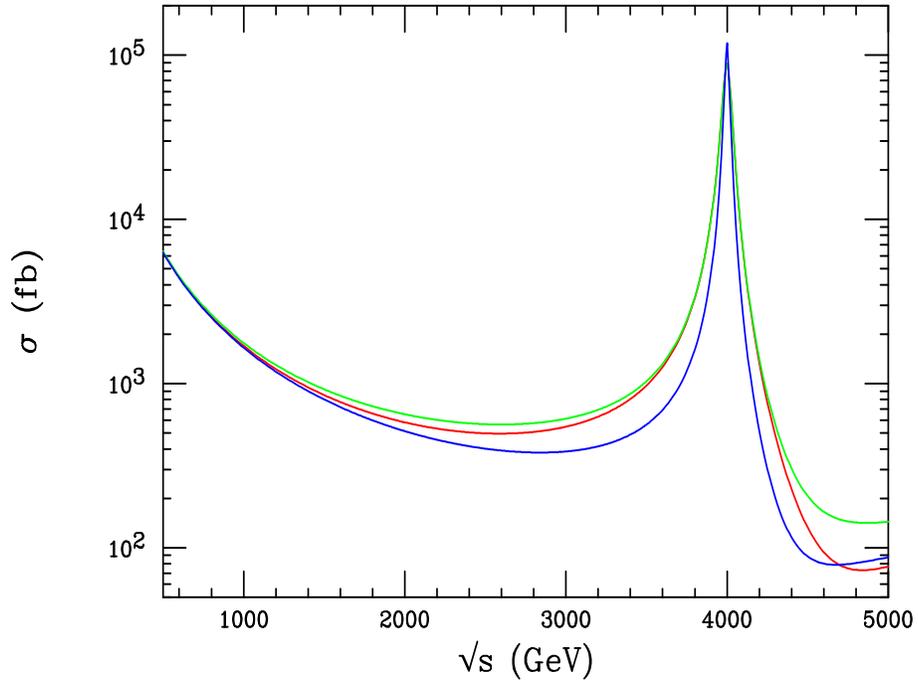,height=9cm,width=12cm,angle=90}}
\vspace*{15mm}
\centerline{
\psfig{figure=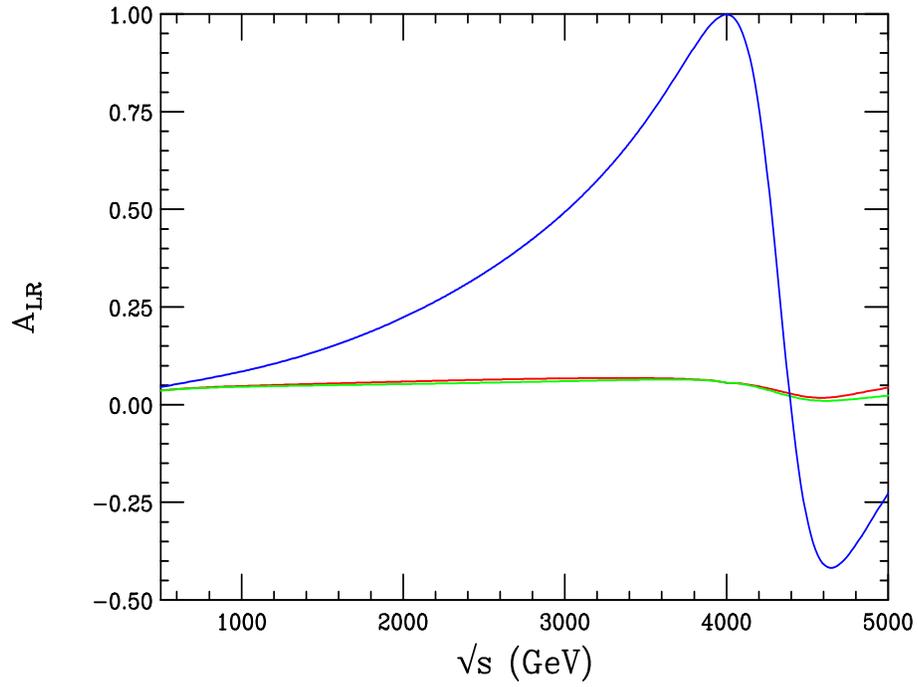,height=9cm,width=12cm,angle=90}}
%\vspace*{-0.9cm}
\caption{Cross section and $A_{LR}$ for Bhabha scattering as discussed in the 
text. The red(green,blue) 
curve is for case I with $x_L=1,x_e=0$($x_L=x_e=0$, case II with 
$x_L=0,x_e=0.5$).}
\label{funny}
\end{figure}
\vspace*{0.4mm}

It is instructive to conveniently bin the contributions to the above sum so 
they can be individually examined. First, we know that the pure SM terms are 
independent of the $x_i$. Next, let us consider the sum of the 
contributions from a single gauge boson KK level due to a specific type of 
gauge boson, \eg, $\gamma^{\pm (n)}$ for fixed $n$. In case I, summing over 
the $Z_2$ even and odd fields yields an effective rescaling
\begin{eqnarray}
(\ell_i\ell_j)^2 &\to& (\ell_i\ell_j)^2 (c_L^2+s_L^2)^2\,, \nonumber \\
(r_ir_j)^2 &\to& (r_ir_j)^2 (c_R^2+s_R^2)^2\,, \nonumber \\
(\ell_i\ell_j)(r_ir_j) &\to& (\ell_i\ell_j)(r_ir_j)(c_Lc_R+s_Ls_R)^2\,.
\end{eqnarray}
For the first two coupling combinations, the resulting scale factor is just 
unity while in the last combination we obtain a factor of 
$\cos^2 n\pi (x_L-x_e)$. This shows that the pure KK tower terms are either 
independent of the $x_i$ or will only depend upon the absolute value of the 
difference $x_L-x_e$ since cosine is an even function. 
In case II the corresponding rescaling factors are $c_L^4, c_R^4$ and 
$c_L^2c_R^2$ thus showing the separate dependence on both $x_L$ and $x_e$. 
The last set of terms are those due to interference 
between, \eg, the SM $\gamma$ and 
$\gamma^{\pm (n)}$ for fixed $n$, again summing over $\pm$ states. This set 
of terms scale similarly to the pure KK contribution except that they are not 
squared. Putting this all together demonstrates that for case I Bhabha 
scattering will only probe the quantity $|x_L-x_e|$ while for case II it will 
probe $x_L$ and $x_e$ independently.

\vspace*{-0.5cm}
\nn
\begin{figure}[htbp]
\centerline{
\psfig{figure=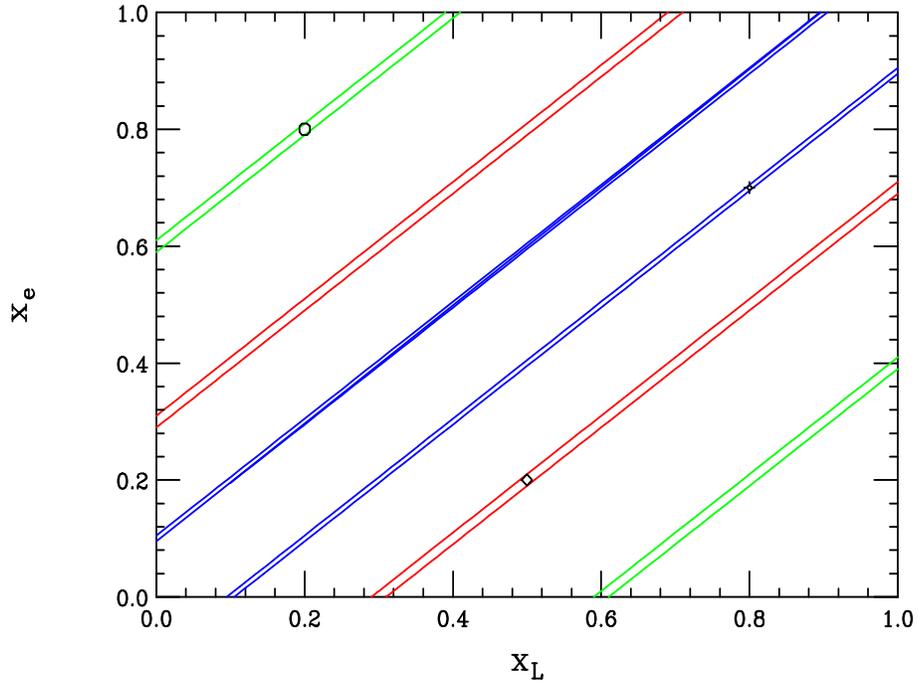,height=9cm,width=12cm,angle=90}}
\vspace*{0.5cm}
\caption[*]{2-parameter fit 
to the values of $x_L$ and $x_e$ from Bhabha scattering data 
in case I as discussed in the text. The circle, plus sign and diamond mark the 
values of the input parameters while the two stripe-like 
pairs of red, green and blue lines 
show the $95\%$ CL allowed regions between them for each input choice. }
\label{thing}
\end{figure}
\vspace*{0.4mm}

To get a feeling for these cross sections we show in Fig.3 the integrated 
cross section and the Left-Right Polarization asymmetry, $A_{LR}$ as a 
function of energy assuming $M=4$ TeV for the common degenerate masses of 
$\gamma^{\pm (1)}$ and $Z^{\pm (1)}$, for three different models. In all 
cases we assume an angular cut $-0.985\leq z=\cos \theta \leq 0.8$; the 
stronger cut for positive $z$ is to remove a significant amount of the SM 
photon pole. In this plot and in the rest of the analysis below neither 
initial state radiation nor beamstrahlung corrections have been applied; for 
more realistic simulations these will need to be included but we know that 
these corrections will not change our results qualitatively. 
To generate the curves in the resonance region, 
family independence was assumed along with the specific values  
$x_Q=x_u=x_d=0.5$. 

How well can we extract $x_{L,e}$ information from Bhabha scattering? Clearly 
at low energies below $\sim 1$ TeV neither the cross section nor $A_{LR}$ 
appears to have any analyzing power. Once the 4 TeV pole is approached, say 
within 
a few widths, we become sensitive to the fact that we have assumed generation 
independence and the particular values above 
for all the other $x_i$. Clearly the 
mass region near the pole should not be included in this fit. Thus we propose 
slicing up the region $1000 \leq \sqrt s \leq 3400$ GeV into 25 points and 
accumulating 100 fb$^{-1}$ of luminosity at each point, determining at each 
both 
$\sigma$ and $A_{LR}$ as well as the Forward-Backward asymmetry, $A_{FB}$, and 
the Polarized Forward-Backward asymmetry, $A_{FB}^{pol}$. (This approach is 
certainly {\it not} optimized but will give us a flavor of 
what we may expect.) We  now employ 
symmetrized cuts $-0.985\leq z \leq 0.985$, assume a 
single beam polarization, $P$, of 
$80\%$, with an error of $\delta P/P=0.003$ and we will also assume an 
error of $0.3\%$ on the integrated 
luminosity. For various input choices of $x_{L,e}$ we can ask what the 
extracted allowed region may be; as typical examples to cover the whole 
parameter space we will assume that $(x_L,x_e)$=
(0.2,0.8), (0.5,0.2), or (0.8,0.7). Fig.4 shows the results of performing 
this fit for case I while Fig.5 shows the corresponding results for case II. 
In case I, since we have already demonstrated that only the 
absolute value of the difference between $x_L$ and $x_e$ can be determined, we
should not be surprised to see that that allowed regions take the form of pairs 
of stripe-like 
narrow bands. For example, when $x_L=0.2$ and $x_e=0.8$, the resulting 
allowed regions should 
surround the lines $x_L=x_e\pm 0.6$ and this is indeed what we 
find. The allowed regions in each of the 
examples lie between the narrowly separated 
pairs of lines; the typical separation of these two lines in the vertical 
direction is $\simeq 0.01$ but varies somewhat from case to case by as much 
as a factor of two. This is the best that one can do for case I using 
data from Bhabha scattering alone. For case II the $x_{L,e}$ can be 
individually determined quite precisely without ambiguity for all three 
examples as shown in Fig.5. One sees that the size of the allowed regions are 
about the same or smaller than the the symbol marking the input values 
themselves.

\vspace*{-0.5cm}
\nn
\begin{figure}[htbp]
\centerline{
\psfig{figure=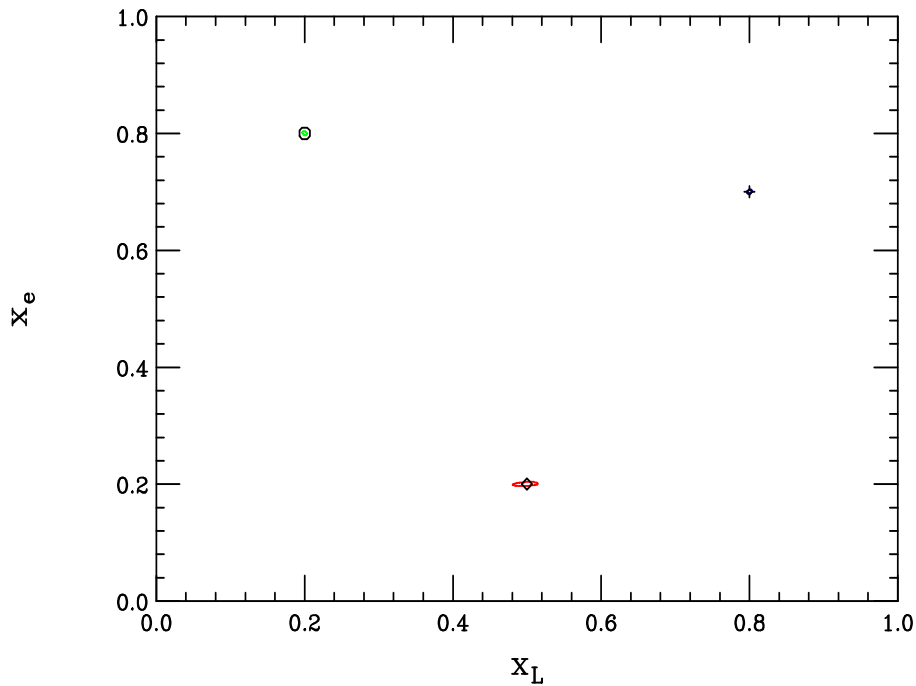,height=9cm,width=12cm,angle=90}}
\vspace*{0.5cm}
\caption[*]{Same as the previous Figure but now for case II; the size of the 
allowed regions are clearly very small in this case and show no ambiguities.}
\label{thing2}
\end{figure}
\vspace*{0.4mm}

\subsection{Fermion Pair Production}

To go further with our analysis 
we need turn our attention to the $s$-channel fermion pair 
production process $e^+e^- \to \bar f f$ which has a cross section given by 
\begin{eqnarray}
{d\sigma \over {d\cos \theta}} &=& {2\pi \alpha \over {s}} \sum_{i,j}
P_{ij}^{ss}\Bigg[{u^2\over {s^2}}\bigg([LL](\ell_i\ell_j)_e(\ell_i\ell_j)_f+
[RR](r_ir_j)_e(r_ir_j)_f\bigg)\nonumber \\
&+&{t^2\over {s^2}}\bigg([LL](\ell_i\ell_j)_e(r_ir_j)_f+
[RR](r_ir_j)_e(\ell_i\ell_j)_f\bigg)\Bigg]\,,
\end{eqnarray}
using the same notation as above. Fig.6 shows that, indeed, the cross section 
and $A_{LR}$ are quite sensitive to variations in the $x_i$ in the same 
$\sqrt s$ region employed in the case of the Bhabha scattering data fit.

\vspace*{-0.5cm}
\nn
\begin{figure}[htbp]
\centerline{
\psfig{figure=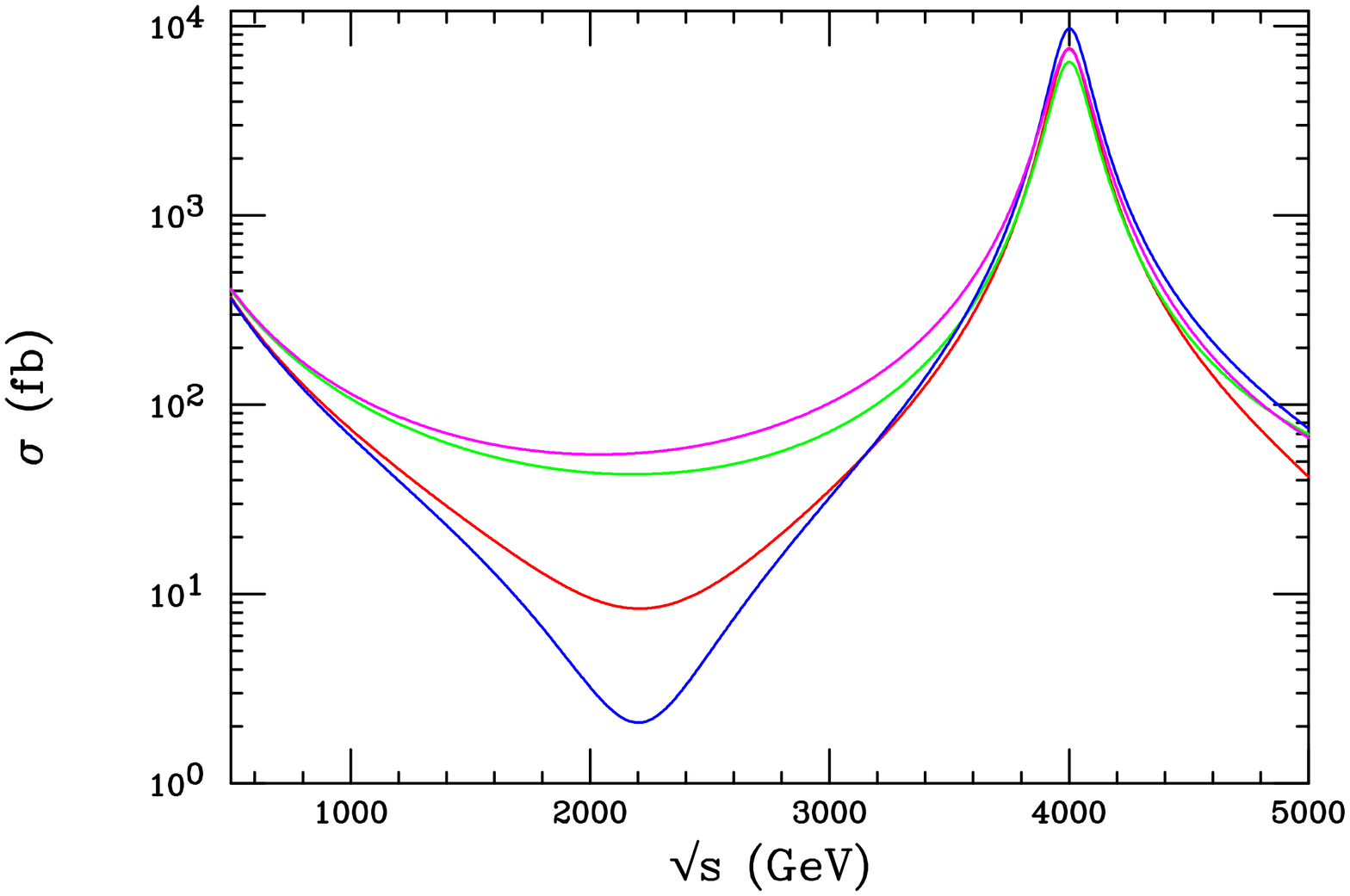,height=9cm,width=12cm,angle=90}}
\vspace*{15mm}
\centerline{
\psfig{figure=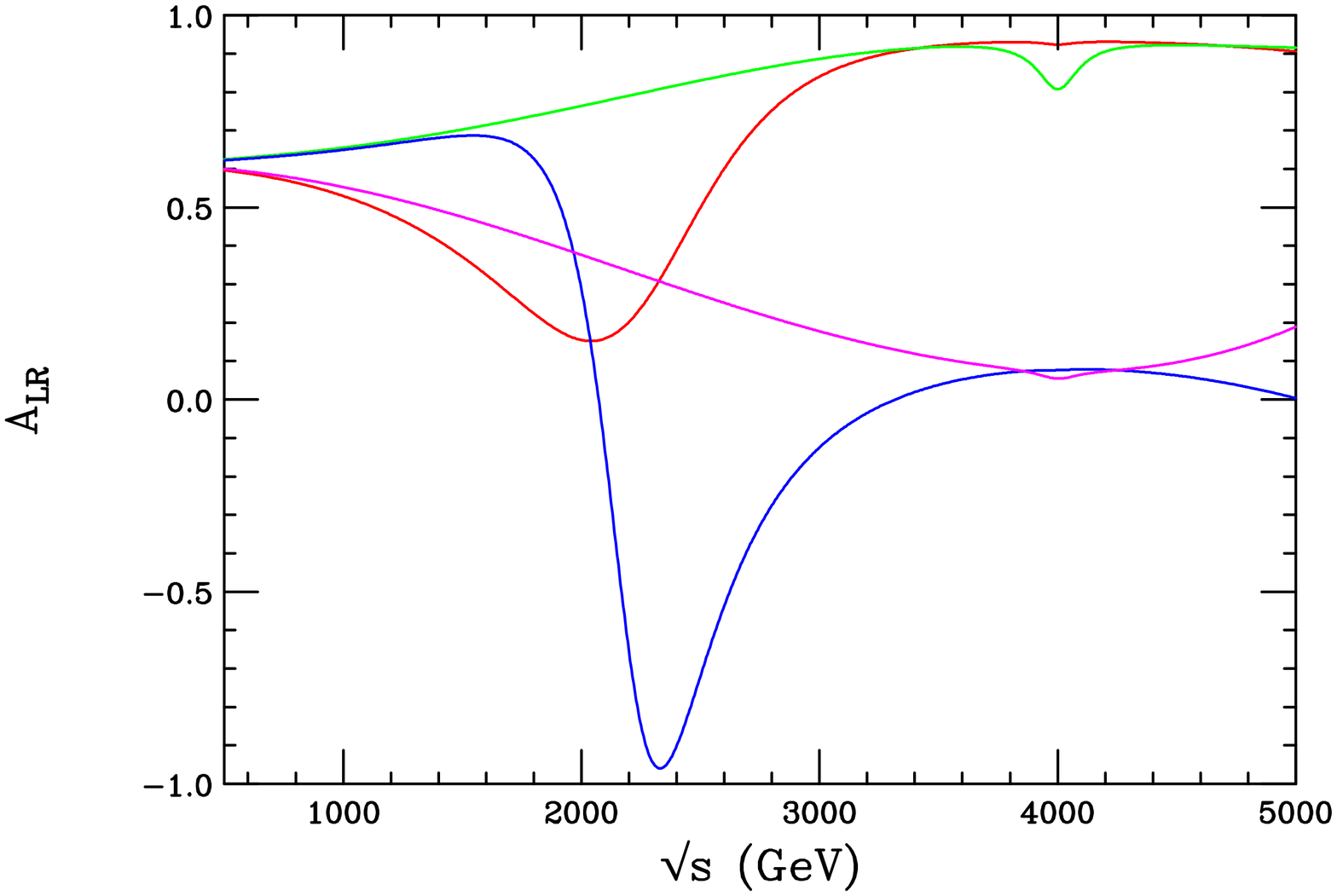,height=9cm,width=12cm,angle=90}}
%\vspace*{-0.9cm}
\caption{Cross section and $A_{LR}$ for $e^+e^-\to \bar b b$ for case I. 
Here we assume 
generation independence along with $x_u=0.01$. The values of $x_L$, $x_e$, 
$x_Q$ and $x_d$ presented in the plot are (0,0,0,0)(red), (0,0,0,1,1)(green), 
(0,1,0,1)(blue) and (0,1,1,0)(purple).}
\label{funnyb}
\end{figure}
\vspace*{0.4mm}

By examining the various pieces that contribute to the cross section as above 
we can learn the structure of its sensitivity to the $x_i$. 
If we concentrate on the pure KK contributions we obtain the 
effective rescalings
\begin{eqnarray}
(\ell_i\ell_j)_e(\ell_i\ell_j)_f &\to& (\ell_i\ell_j)_e(\ell_i\ell_j)_f
(c_L^ec_L^f+s_L^es_L^f)^2\,, \nonumber \\
(r_ir_j)_e(r_ir_j)_f &\to& (r_ir_j)_e(r_ir_j)_f(c_R^ec_R^f+s_R^es_L^f)^
2\,, \nonumber \\
(\ell_i\ell_j)_e(r_ir_j)_f &\to& (\ell_i\ell_j)_e(r_ir_j)_f
(c_L^ec_R^f+s_L^es_R^f)^2\,, \nonumber \\
(r_ir_j)_e(\ell_i\ell_j)_f &\to& (r_ir_j)_e(\ell_i\ell_j)_f
(c_R^ec_L^f+s_R^es_L^f)^2\,,
\end{eqnarray}
which for case I means that at least this part of the cross section depends 
only upon the 
absolute values of four distinct differences that one can form between 
the various $x_i$. The SM-KK tower interference terms are found to scale 
exactly as above without squaring the factors on the right 
making this conclusion 
valid for almost the entire cross section. Note that unlike the case of Bhabha 
scattering, there are no terms in the cross section which are $x_i$ 
independent beyond those arising from the SM. For case II the same expressions 
as above are found to hold but with all the terms proportional to 
$s^{e,f}_{L,R}$ set to zero.

\vspace*{-0.5cm}
\nn
\begin{figure}[htbp]
\centerline{
\psfig{figure=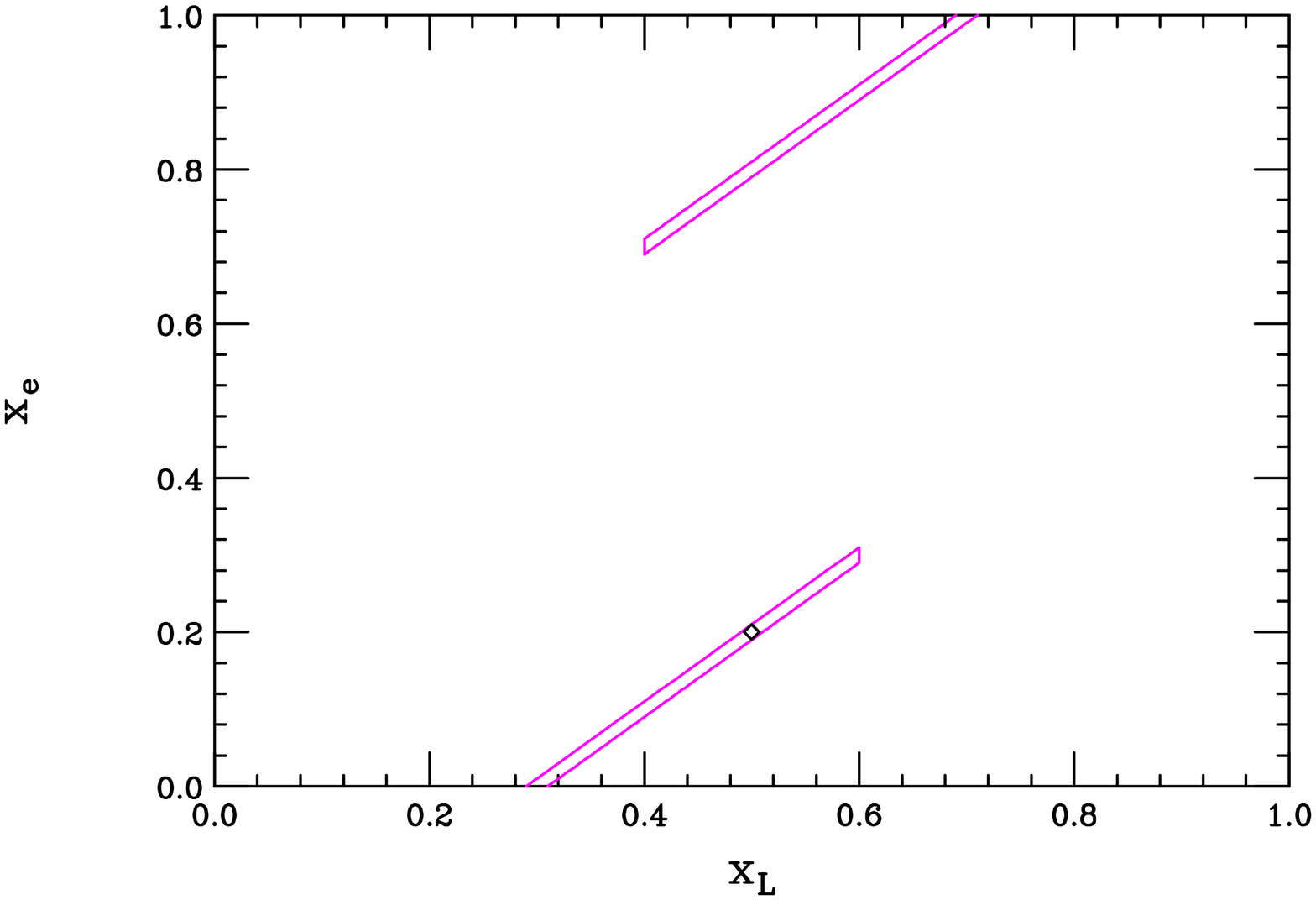,height=9cm,width=12cm,angle=90}}
\vspace*{15mm}
\centerline{
\psfig{figure=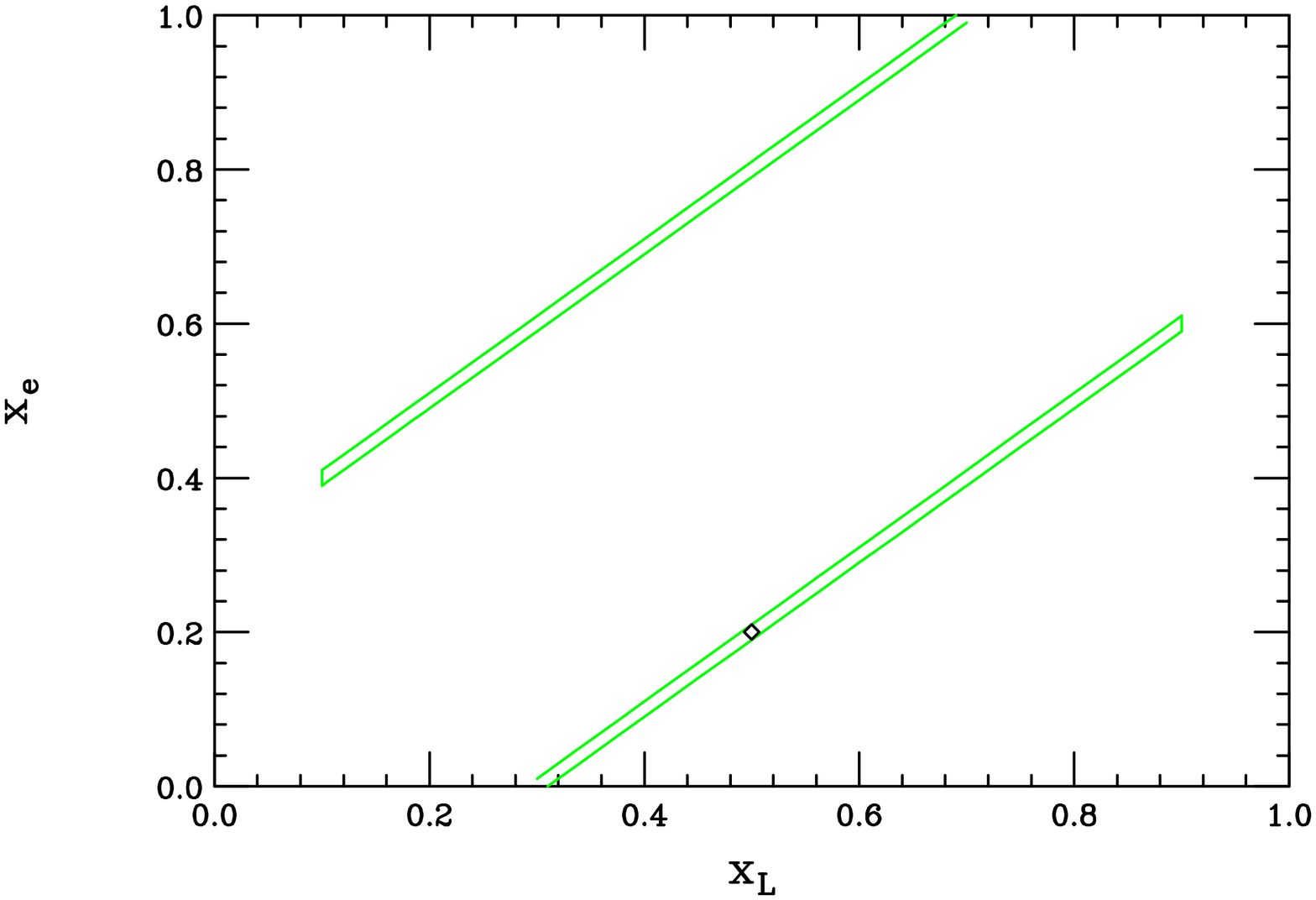,height=9cm,width=12cm,angle=90}}
%\vspace*{-0.9cm}
\caption{$95\%$ CL combined 4-parameter fit to Bhabha scattering and 
$\bar bb$ production projected onto the $x_e-x_L$ plane in case I as 
discussed in the text. In the top(bottom) 
panel $x_d=0.6(0.4)$ and $x_Q=0.9(0.1)$ have been assumed.}
\label{funn}
\end{figure}
\vspace*{0.4mm}

One might believe that since the cross section for case I now depends 
upon four absolute values of $x_i$ differences while Bhabha scattering 
provides a fifth absolute difference, 
it may be possible to uniquely determine all the $x_i$'s 
when the two sets of data are combined. Unfortunately, this is not the case as 
a simple example shows. Imagine the $x_i$ take on the particular set of 
values $x_i^0$; 
certainly if one adds or subtracts a common value, $\delta$,  to all of the 
$x_i^0$ the 
absolute values of the differences upon which the cross sections depend 
will be left invariant as long as none of the $x_i$ exceed unity or is less 
than zero as a result of this shift. This implies that the size and location 
of the $95\%$ CL allowed region on the $x_L-x_e$ projection for case I 
will critically 
depend, \eg, in the case of $b$-quarks, on both $x_Q$ and $x_d$. An example of 
this is shown in Figs.7 and 8 where Bhabha scattering and $e^+e^- \to \bar bb$ 
data are combined in an effort to extract the four parameters $x_L$, $x_e$, 
$x_Q$ and $x_d$. (Here a b-tagging efficiency of $60\%$ has been used and QCD 
corrections have been included; we fit to the data taken at the same energy 
points with the same integrated luminosity as we have already done for the 
Bhabha analysis.)

\vspace*{-0.5cm}
\nn
\begin{figure}[htbp]
\centerline{
\psfig{figure=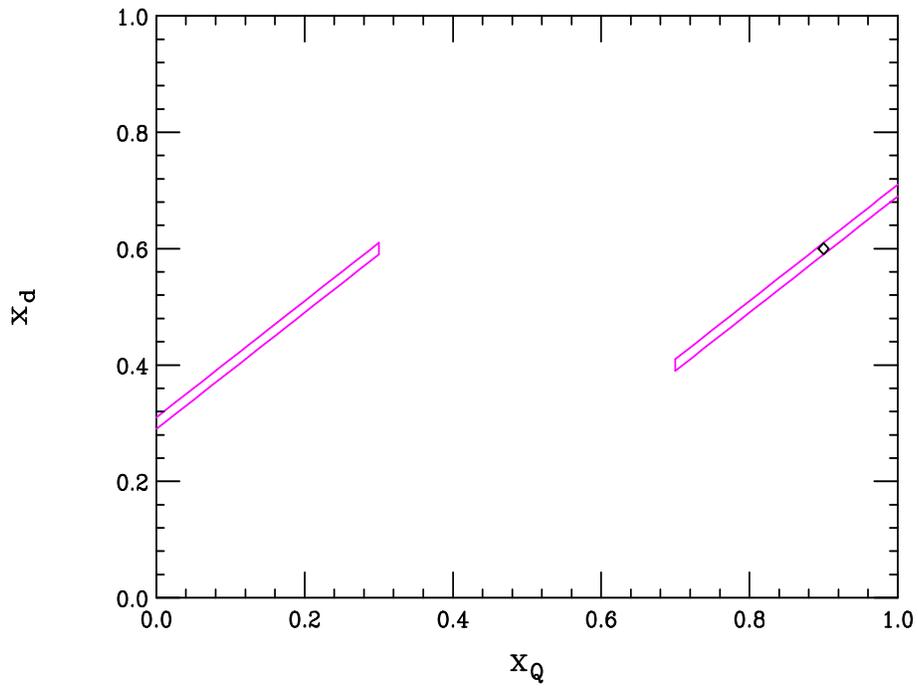,height=9cm,width=12cm,angle=90}}
\vspace*{15mm}
\centerline{
\psfig{figure=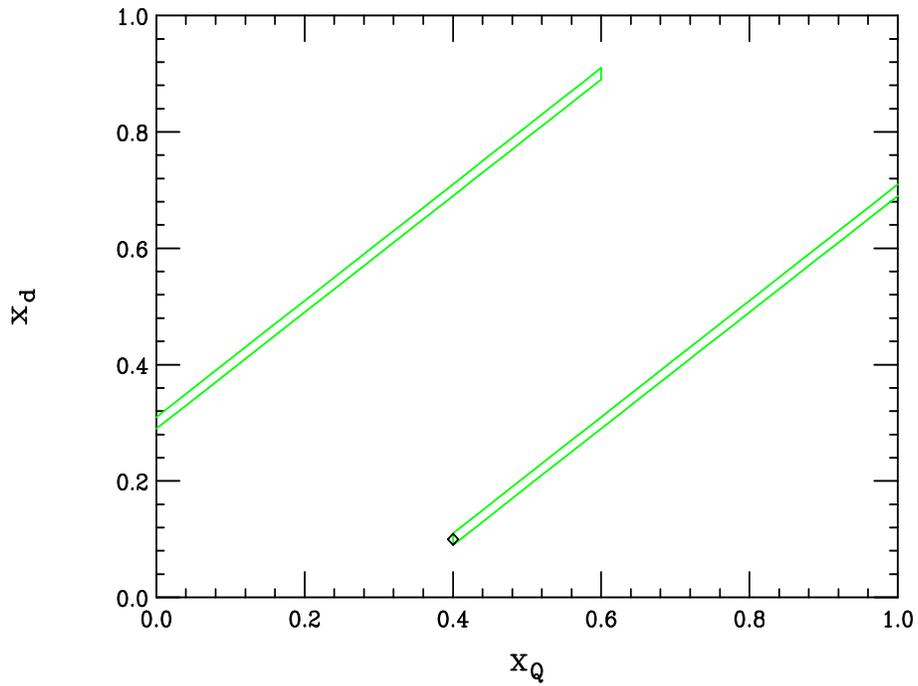,height=9cm,width=12cm,angle=90}}
%\vspace*{-0.9cm}
\caption{Companion plots to the previous figures for the $95\%$ CL 4-parameter 
fits projected down into the $x_d-x_Q$ plane.}
\label{funnybunny2}
\end{figure}
\vspace*{0.4mm}

As we can see from these figures the inclusion of $e^+e^- \to \bar bb$ data 
does decrease the size of the stripe-shaped 
allowed range extracted for $x_{e,L}$ from 
Bhabha scattering but not by a very great amount. 
We also see that the locations of 
the allowed regions do depend on the assumed values of $x_{d,Q}$. We should 
be reminded that since these plots are 2-dimensional projections of the 
4-dimensional fits, the various points in the allowed regions shown in Figs. 
7 and 8 are highly correlated.

In case II there are no ambiguities in any of the couplings and, as in the case 
of Bhabha scattering, only a very small allowed region is found from the fit. 
The size of the allowed region on the $x_e-x_L$ plane is this case is 
found to be independent of the values of $x_{d,Q}$ and not much different 
than that obtained from Bhabha scattering alone. 
All three examples are compared in Fig.9.

\vspace*{-0.5cm}
\nn
\begin{figure}[htbp]
\centerline{
\psfig{figure=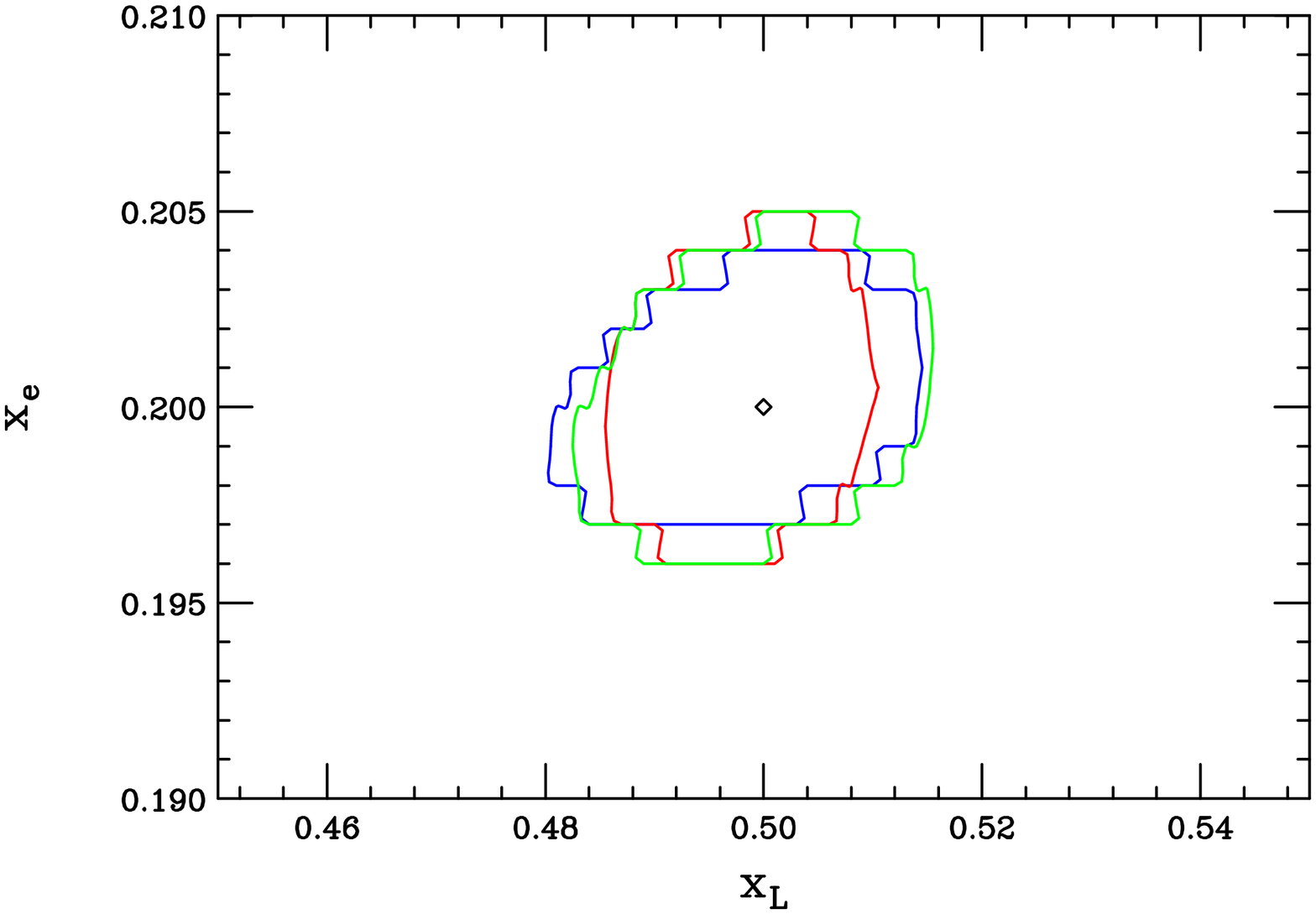,height=9cm,width=12cm,angle=90}}
\vspace*{0.5cm}
\caption[*]{$95\%$ CL fits for case II. The blue(red,green) curve corresponds 
to a fit of Bhabha scattering data alone(the 4-parameter fit with $x_Q=0.9$ 
and $x_d=0.6$, the 4-parameter fit with $x_Q=0.1$ and $x_d=0.4$).}
\label{thing22}
\end{figure}
\vspace*{0.4mm}

From this analysis it is clear that for case II precision measurements of the 
various fermion SM locations  is relatively easy to achieve and is quite 
probably dominated by systematic uncertainties. For case I precision 
measurements of fermion locations 
may be possible as well although it would seem that fits to only 
one or two processes leads to bounded, striped-shape regions. Clearly as 
more and more final state fermion locations are probed the lengths and 
locations of these stipes will be reduced. It may be possible that if data on 
a sufficient number of final states can be collected the allowed 
striped-shaped regions will be cut down to a sufficient size so that a precise 
location determination is achievable. Such an analysis is, however, beyond the 
scope of the present paper so that we must for the moment 
be satisfied with our rather narrow striped allowed regions for the fermions' 
locations.

\subsection{On-Resonance Observables}

In our earlier work{\cite {me}} it was shown that once we determine that the 
spin of a resonance in lepton-pair collisions is one (from decay angular 
distributions) it will be straightforward to determine whether or not it is 
a KK state or a more typical $Z'$ provided beam polarization is available. 
This rests on the use of factorization 
tests, \ie, linear relationships between measurements of observables made atop 
the resonance. As is well-known, on a $Z'$ resonance the tree-level coupling 
factorization relationship 
$A_{FB}=A_{LR}A_{FB}^{pol}$ is known to hold for any final state fermion $f$. 
However, in the KK case, a single particle resonance is not what is produced. 
In case I(II) a coherent, resonating combination of 4(2) states is produced 
simultaneously so we expect relations such as the above to fail. (In earlier 
work{\cite {me}} this was explicitly shown to happen for the rather simple 
case II so that it is rather obvious that it also happens for more complex 
case I. In this discussion we omit the possibility of any tiny 
mass splittings of order 1 GeV between the $Z$ and $\gamma$ states.) 
This can clearly be seen as follows:   
a short exercise shows that the above observables on resonance can be written 
as 
\begin{eqnarray}
A_{FB}^f &=& {3\over 4} {A_1\over D}\nonumber \\
A_{FB}^{pol}(f) &=& {3\over 4} {A_2\over D}\nonumber \\
A_{LR}^f &=& {A_3\over D}\,,
\end{eqnarray}
where $f$ labels the final state fermion and we have defined the coupling 
combinations
\begin{eqnarray}
D &=& \sum_{i,j}\Gamma_i^{-1}\Gamma_j^{-1}(\ell_i\ell_j+r_ir_j)_e
(\ell_i\ell_j+r_ir_j)_f\nonumber \\
A_1 &=& \sum_{i,j}\Gamma_i^{-1}\Gamma_j^{-1}(\ell_i\ell_j-r_ir_j)_e
(\ell_i\ell_j-r_ir_j)_f\nonumber \\
A_2 &=& \sum_{i,j}\Gamma_i^{-1}\Gamma_j^{-1}(\ell_i\ell_j+r_ir_j)_e
(\ell_i\ell_j-r_ir_j)_f\nonumber \\
A_3 &=& \sum_{i,j}\Gamma_i^{-1}\Gamma_j^{-1}(\ell_i\ell_j-r_ir_j)_e
(\ell_i\ell_j+r_ir_j)_f\,,
\end{eqnarray}
with $\Gamma_i$ the individual widths of the contributing KK states. For case 
I, the sum extends over $\gamma^{\pm (1)}$ and $Z^{\pm (1)}$ while for case II 
only the states $\gamma^{+(1)}$ and $Z^{+(1)}$ are included. Note that as 
long as all of the $\Gamma_i$ are of comparable magnitude factorization 
relations of the type above are impossible to satisfy. 
Only in the limit where one of the 
$\Gamma_i$ dominates the sum (or, of course, when only one state is produced) 
will factorization be obtainable. Once factorization fails we then know that we 
are producing more than a simple single resonance. 

We note in passing that it is often asked whether it 
may also be possible to distinguish a single $Z'$-like resonance from a KK 
state by the shape of the excitation spectrum. Such an analysis would need to 
include the effects of both the slight mass difference in the $\gamma$ and 
$Z$ states, initial state radiation, finite mass resolution, 
as well as the appropriate 
beamstrahlung spectrum corrections and has yet to be performed. Fig.10 shows 
what a comparison of a KK excitation and a single fitted Breit-Wigner(BW) may 
look like after ISR and beamstrahlung have been de-convoluted. Near the peak 
the BW is somewhat more narrow than the actual distribution and overestimates 
the cross sections in the tails. The actual KK distribution is also seen to be 
somewhat narrower above than below the resonance peak. Given these small 
differences it is clear that a detailed analysis 
must be performed to demonstrate 
whether or not a KK state can be distinguished from a single BW due to the 
importance of systematic uncertainties.

\vspace*{-0.5cm}
\nn
\begin{figure}[htbp]
\centerline{
\psfig{figure=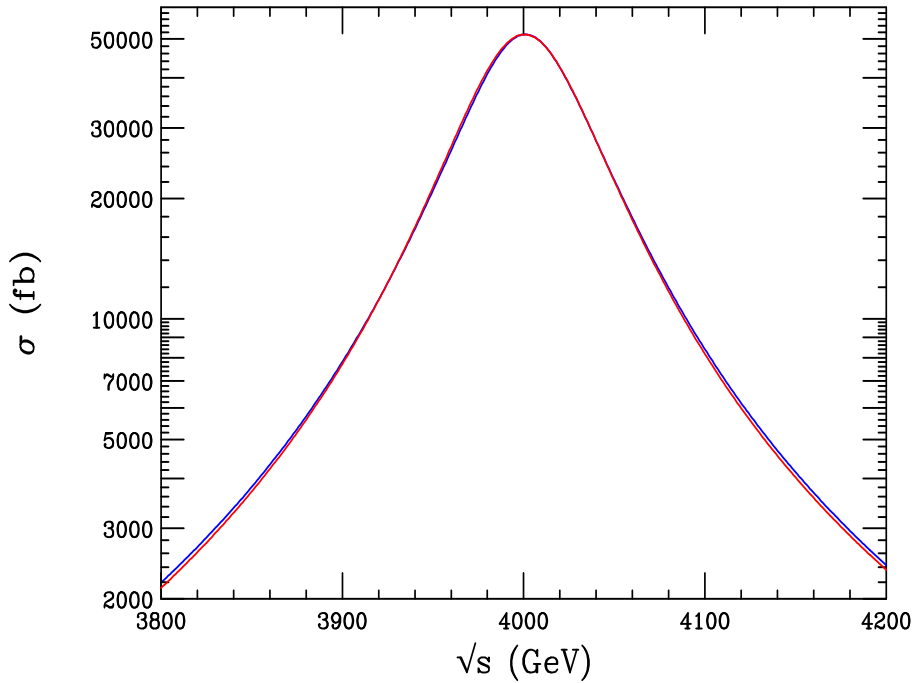,height=9cm,width=12cm,angle=90}}
\vspace*{0.5cm}
\caption[*]{Comparison of a single fitted Breit-Wigner(blue) and a case I KK 
excitation(red) with generation independent couplings in 
$e^+e^- \to \mu^+\mu^-$. 
The values $x_L=x_u=0.5$, $x_e=0.2$, $x_d=0.6$ and $x_Q=0.9$ have been 
assumed. No ISR or beamstrahlung corrections have been applied; note that the 
BW tends to overestimate the cross section when more than $\sim 100$ GeV 
from the resonance in this case.}
\label{thing33}
\end{figure}
\vspace*{0.4mm}

With the above discussion, we may ask what information can be obtained on the 
$x_i$ from resonance measurements; certainly statistics is not problem with 
more than $\sim 10^{6-7}$ events expected per ab$^{-1}$. 
With these huge statistics it is clear that any such a determination 
of couplings will be systematics limited for 
either case I or case II. The simplest case to 
analyze, in the absence of any family symmetry and in analogy with Bhabha 
scattering, is the $e^+e^-$ final state. In this case, the couplings in Eqs. 
(15) and (16) only depend upon the two unknowns $x_{e,L}$; however, the 
asymmetry observables also depend on the three(one) independent ratios of 
widths, $\Gamma_i/\Gamma_j=R_{ij}$, in case I(II). 
Thus for case I, there are not enough 
observables in this sector alone to fix all of the parameters while a unique 
solution seems to be obtainable in case II.  For case I it is clear that one 
must combine the data from several flavor sectors in order to obtain a 
sufficient number of constraints to determine the relevant $R_{ij}$ and $x_i$ 
from a rather large simultaneous fit. Such an analysis is beyond the scope of 
this paper. 
For case II, knowing $x_{e,L}$, and the single width ratio $R_{12}$, it 
would appear that we can proceed to all the other sectors and determine the 
corresponding $x_i$ without much difficulty. However in the case of electrons 
there are actually only two independent asymmetries since $A_2=A_3$. Thus, as 
in case I, data from other final states will need to be included and a large 
simultaneous fit performed. It is clear from this discussion that it will be 
somewhat easier to isolate the positions of the various SM fermions in the 
extra dimensions by using off-resonance measurements than via on-peak data.

\section{Summary and Conclusions}

In this paper we have investigated the capability of future lepton colliders 
to probe the locations of the SM fermions when they are `stuck' at specific 
points in extra dimensions as suggested by the Arkani-Hamed-Schmaltz model. 
There were several major steps necessary to perform this analysis which we now 
summarize.

\begin{itemize}

\item  We consider two classes of models: I and II; they differ by whether or 
not the $Z_2$ odd components of the gauge KK towers are allowed to propagate 
and couple to the SM fermions. These two classes of models are identical when 
the fermions live at the fixed points put differ significantly if the fermions 
can live at arbitrary locations. 

\item  In order to investigate the fermion locations at an accelerator the 
first requirement is to determine the scale of the the compactification which 
determines the mass of the KK gauge boson spectrum. Direct searches for such 
states at the Tevatron only provide limits{\cite {me}} 
on the $\sim 0.8$ TeV range while 
those from indirect searches, \ie, the use of precision electroweak data, 
can provide limits in the several TeV range. For the case at hand, allowing 
for the left-handed electron and $\mu$ to occupy different locations 
within the thick wall, we found KK mass constraints that are fairly sensitive 
to the separation of these two fields as well as the SM Higgs mass. Scanning 
over the parameter space a typical bound we obtained by this procedure is 
$\sim 4$ TeV. In some cases it may be possible to obtain a higher average 
bound by considering various FCNC processes but these are much more model 
dependent. Clearly, to reach and probe the $\sim 4$ TeV scale with sufficient 
statistics will require CLIC or a 
Muon Collider with very high integrated luminosities $\sim 1~ab^{-1}$. 

\item  Since this center of mass energy is quite high it is important that 
we first ascertain 
what can be learned from data taken below the resonance. In this energy range, 
for any given process such as $e^+e^- \to f\bar f$, the cross section and 
various asymmetries depend {\it only} upon the locations of these specific 
fermions, \ie, $e_{L,R}$ and $f_{L,R}$. The simplest 
process to consider is thus 
Bhabha scattering. Dividing the mass range up into bins of equal integrated 
luminosity and making use of the cross section and asymmetry observables we 
were able to obtain the allowed regions for the locations of $e_{L,R}$ which 
differed significantly for cases I and II. In case I we found, consistent 
with theoretical expectations, that only the quantity 
$|x_L-x_e|$ could be precisely 
determined, with the allowed parameter space thus appearing as striped-shaped 
regions in the $x_L-x_e$ plane, 
whereas in case II both $x_{L,e}$ can be well-determined separately. 
Adding information from another process, such as $e^+e^- \to b\bar b$,  
led to a four parameter fit; for case I the size and locations of 
the allowed 
regions in location space were shown to be rather sensitive to the assumed 
values of the input parameters. On the otherhand, in 
case II the fermion location sensitivity was observed not 
to be sensitive to the choice of input. If case II holds, it is 
clear that from a generalization of our procedure to other fermion final 
states that locating the SM fermions will be rather straightforward using data 
obtained below the first KK pole. For case I such a situation is more 
problematic since the allowed regions, projected in two dimensions, are 
stripe-shaped. It may be, however, that if enough final states are accessible, 
a large simultaneous fit can lead to rather small allowed regions for the 
various fermion locations.

\item  We demonstrated that on-resonance measurements will be much harder to 
use in determining fermion locations. The reason for this is that for any 
particular fermion decay mode, the locations of {\it all} the SM fermions are 
involved in determining the value of the associated observables. Fortunately, 
we demonstrated that the these additional dependencies can be isolated into a 
only one(for case II) or three(for case I) additional 
unknowns when conducting the 
appropriate fits. Even with this simplification, however, we argued that large 
simultaneous fits involving many observables will be necessary to extract 
location information for either case I or II in order to obtain a sufficient 
number of independent quantities to fit the rather large number of unknowns. 
Given the huge on-resonance statistics, the errors in such fits will certainly 
be systematics dominated. In addition, 
we speculated on the possibility of distinguishing a KK resonance from a single 
Breit-Wigner through its' shape; such an analysis will require a detailed 
study including initial state radiation, beamstrahlung and finite energy 
resolution effects and is beyond the scope of this paper. 

\end{itemize}

From our analysis it is clear that future lepton colliders will provide a 
means to map out the positions of SM fermions if they are localized in extra 
dimensions.

\noindent{\Large\bf Acknowledgements}

The author would like to thank N. Arkani-Hamed, Y. Grossman and M. Schmaltz 
for discussions on localized fermion models during the early stages of this 
work. The author would also like to thank H. Davoudiasl and J.L. Hewett for 
general discussions on models with extra dimensions.

\newpage

%
%%%%%%%%%%%%%%%%%%--- References
%%%%%%%%%%%%%%%%%%%%%%%%%%%%%%%%%%%%%%%%%%%%%%%%%%%%%%%
\def\MPL #1 #2 #3 {Mod. Phys. Lett. {\bf#1},\ #2 (#3)}
\def\NPB #1 #2 #3 {Nucl. Phys. {\bf#1},\ #2 (#3)}
\def\PLB #1 #2 #3 {Phys. Lett. {\bf#1},\ #2 (#3)}
\def\PR #1 #2 #3 {Phys. Rep. {\bf#1},\ #2 (#3)}
\def\PRD #1 #2 #3 {Phys. Rev. {\bf#1},\ #2 (#3)}
\def\PRL #1 #2 #3 {Phys. Rev. Lett. {\bf#1},\ #2 (#3)}
\def\RMP #1 #2 #3 {Rev. Mod. Phys. {\bf#1},\ #2 (#3)}
\def\NIM #1 #2 #3 {Nuc. Inst. Meth. {\bf#1},\ #2 (#3)}
\def\ZPC #1 #2 #3 {Z. Phys. {\bf#1},\ #2 (#3)}
\def\EJPC #1 #2 #3 {E. Phys. J. {\bf#1},\ #2 (#3)}
\def\IJMP #1 #2 #3 {Int. J. Mod. Phys. {\bf#1},\ #2 (#3)}
\def\JHEP #1 #2 #3 {J. High En. Phys. {\bf#1},\ #2 (#3)}

\end{document}